\documentclass{aa} 

\usepackage{graphicx}      
\usepackage{txfonts}       
\usepackage{amsmath}       
\usepackage{natbib}        
\bibpunct{(}{)}{;}{a}{}{,} 
\usepackage{algorithm}
\usepackage{algpseudocode}

\usepackage[colorlinks=true, allcolors=blue]{hyperref}

\def\apjl{ApJ}
\def\apjs{ApJS}
\def\ao{Appl.~Opt.}

\def\aap{A\&A}

\begin{document} 
	
	\title{An adaptive parameter optimization method for astronomical image alignment using Bayesian optimization}
	\subtitle{I. A hierarchical search strategy for FWHM and SNR}
	
	\titlerunning{Adaptive Parameter Optimization for Image Alignment}
	\authorrunning{Y. Zhang \& Y. Zhang}
	
	\author{Yongjie Zhang\inst{1} 
		\and Yigong Zhang\inst{1}\fnmsep\thanks{Corresponding author: kmuzhyg@163.com}
		\and Zhenjun Zhang\inst{2}
		\and Xiangming Cheng \inst{3}
		\and Lei Xiong \inst{1}
		\and Xiaoguang Yu\inst{4}
		\and Jie Su\inst{5} 
		\and Jiancheng Wang\inst{3}
		\and Haoyang Guo\inst{1} }
	\institute{School of Information Engineering, Kunming University, Kunming 650214, China
		\and School of Computer of Science and Information Engineering, Anyang Institute of Technology, Anyang 455000, China
		\and Yunnan Observatories, Chinese Academy of Sciences, Kunming 650216, China
		\and Fujian Polytechnic Normal University, Fuqing 350300, China
		\and College of Big Data, Yunnan Agricultural University, Kunming 650201, China}
	\date{Received [Date]; accepted [Date]}

	\abstract
	{The alignment and stacking of astronomical images are fundamental steps for detecting faint objects and performing high-precision astrometry.}
	{In traditional alignment workflows, the extraction of source lists is critically dependent on key parameters such as the Full Width at Half Maximum (FWHM) and the Signal-to-Noise Ratio (SNR) threshold. These parameters are often selected manually through an inefficient trial-error process that lacks objectivity and does not guarantee optimal results.}
	{We present an adaptive method for optimizing astronomical image alignment parameters based on Bayesian Optimization (BO). We frame the parameter search as an optimization problem, with an objective function designed to maximize the number of successfully matched source pairs. By employing a hierarchical search strategy, we perform an efficient global search for FWHM and SNR to automatically determine the optimal combination for a given observational dataset.}
	{Experimental results demonstrate that our method effectively handles image data with varying seeing conditions and background noise levels. It rapidly converges to a robust set of alignment parameters, achieving sub-pixel accuracy and significantly improving the automation level and success rate of the alignment process.}
	{This work may provide a useful basis for developing large-scale, automated astronomical data processing pipelines.}
	
	
	\maketitle
	\nolinenumbers
	
	\section{Introduction}
	\label{sec:intro}
	
	Modern astronomical optical observations play a significant role in various fields, including star formation and evolution, deep space exploration \citep{Meurisse2020}, and near-Earth object (NEO) defense \citep{Bancelin2012}. The prerequisite for achieving effective superposition is precise image alignment. In the process of alignment, due to the influence of random factors such as atmospheric observation conditions and background noise, the selection of alignment parameters directly affects the success of alignment. Taking the widely used DAOStarFinder algorithm \citep{Stetson1987} as an example, its detection results are mainly affected by two key parameters: FWHM and SNR threshold. FWHM describes the degree of dispersion of the star image and is directly related to the atmospheric seeing conditions of the night; the SNR threshold determines the algorithm's sensitivity to distinguish real source points from background noise.
	
	In actual observations, due to variations in atmospheric seeing, telescope tracking errors, and background fluctuations, the optimal values of these parameters are not constant. Inappropriate settings typically lead to two failure modes: (1) Overly strict parameters result in too few detected sources to establish a reliable geometric transformation; (2) Overly lenient parameters introduce excessive noise-induced spurious sources or cosmic rays, which clutter the source list and confound the matching algorithm, eventually leading to alignment failure or reduced precision.
	
	The traditional methods of parameter adjustment such as manual trial and error are not only inefficient but also highly subjective, making it difficult to ensure that the parameters found are the global optimal solution. Especially under complex observation conditions, fixed processing parameters will lead to a high failure rate of alignment, making it difficult to meet the requirements of automated processing of massive data in large-scale sky surveys.
	
	The alignment and stacking of optical images are foundational techniques for scientific goals ranging from the discovery of faint celestial bodies to high-precision astrometry \citep{Tyson1992, Bernstein2004, Malbet2021,Zhang2021}. These techniques have been extensively applied in the detection of faint moving objects \citep{Li2020, Wang2017}. In recent years, with the assistance of high-speed cameras and GPU parallel computing, these methods have evolved into "Synthetic Tracking" (ST), enabling the detection of extremely faint and fast-moving targets while achieving milliarcsecond-level astrometric precision \citep{Shao2014, Zhai2014, Zhai2018, Zhai2020}. However, whether employing classical image subtraction \citep{Zackay2016} or advanced ST techniques, the internal processing pipelines inevitably involve source detection stages. Consequently, these methods remain constrained by the persistent challenge of optimal parameter configuration.
	
	With the exponential growth of astronomical data, data-driven approaches represented by machine learning and Bayesian statistics are playing an increasingly vital role. Machine learning has become a powerful tool for complex classification and identification tasks. In aerospace engineering, fuzzy neural networks have been employed for autonomous star pattern recognition \citep{Hong2000}. Recently, Convolutional Neural Networks (CNNs) have been applied to tasks such as image denoising \citep{Liu2025}, faint object detection \citep{Chen2025}, and fast identification of spatial targets \citep{E2022}. On a broader scientific level, machine learning is utilized for the precise identification of star cluster members from large catalogs \citep{Chi2025}. These studies leverage powerful non-linear fitting capabilities to learn complex patterns for classification and regression problems.
	
	In parallel, Bayesian methods have achieved significant progress in astrophysical model construction due to their unique advantages in handling uncertainty and performing parameter inference. For instance, Bayesian sampling techniques have been utilized to invert the orbital parameters of exoplanets \citep{Bao2025}. Similar probabilistic approaches have been used to infer the physical states of high-redshift submillimeter galaxies \citep{Yang2019}. These applications demonstrate that Bayesian methods serve as robust scientific modeling tools for exploring the parameter spaces of complex physical systems. Furthermore, in the field of instrument calibration, Bayesian methods have been adopted to improve the precision of photometric calibration \citep{Goodeve2025}.
	
	Despite these advancements, a systematic application of Bayesian Optimization (BO) to the front-end parameter adaptation of image alignment remains largely unexplored. The main focus of this research is to improve the automated optimization technology for processing parameters in the data processing workflow. By developing a method that can automatically and quickly adapt to different observational data and determine the optimal alignment parameters, the efficiency of astronomical data processing has been effectively enhanced. This paper transforms the parameter selection problem in image alignment into an optimization problem and uses the Bayesian Optimization algorithm to solve it. A parameter adaptive framework was designed and implemented, which can automatically explore and determine the optimal combination of FWHM and SNR threshold, significantly improving the success rate and robustness of image registration, and verifying the effectiveness of the parameter adaptive strategy.
	
	This paper is structured as follows. Section 2 describes our methodology in detail, where we reframe the parameter selection for image alignment as a formal optimization problem and introduce the hierarchical Bayesian Optimization (HBO) framework designed to solve it. In Section 3, we present comprehensive experimental results using data from the Yunnan Observatory's 1-meter telescope. These experiments validate our method's ability to handle varying seeing conditions, demonstrate its convergence to sub-pixel astrometric precision, and confirm its efficiency for automated processing. Finally, Section 4 provides a summary of our work and its key conclusions.
	
	\section{Methodology}
	\label{sec:method}
	
	The robustness of the traditional astronomical image registration process is limited by the manual setting of source point detection parameters. Aiming at the limitations of traditional methods, this paper focuses on the adaptive optimization of parameters at the early stage of the alignment process. We introduce a BO algorithm to achieve adaptive tuning of FWHM and SNR thresholds. This framework aims to maximize the number of successfully matched source pairs by constructing a probabilistic surrogate model using Gaussian Processes (GP) and employing an Expected Improvement (EI) acquisition function to balance "exploration" and "exploitation." The successfully matched source pairs are used to calculate the affine transformation matrix, and the more the number of source pairs involved in the calculation, the higher the success rate of calculating the affine transformation matrix.
	
	Directly applying BO to parameters on full-resolution images can be computationally prohibitive, as each evaluation of the objective function—which involves source detection and matching—is time-consuming. To address this challenge, we designed an efficient hierarchical search strategy that synergistically balances computational speed with optimization accuracy.
	
\begin{figure*}[t!] 
		\centering
		\includegraphics[width=0.9\textwidth]{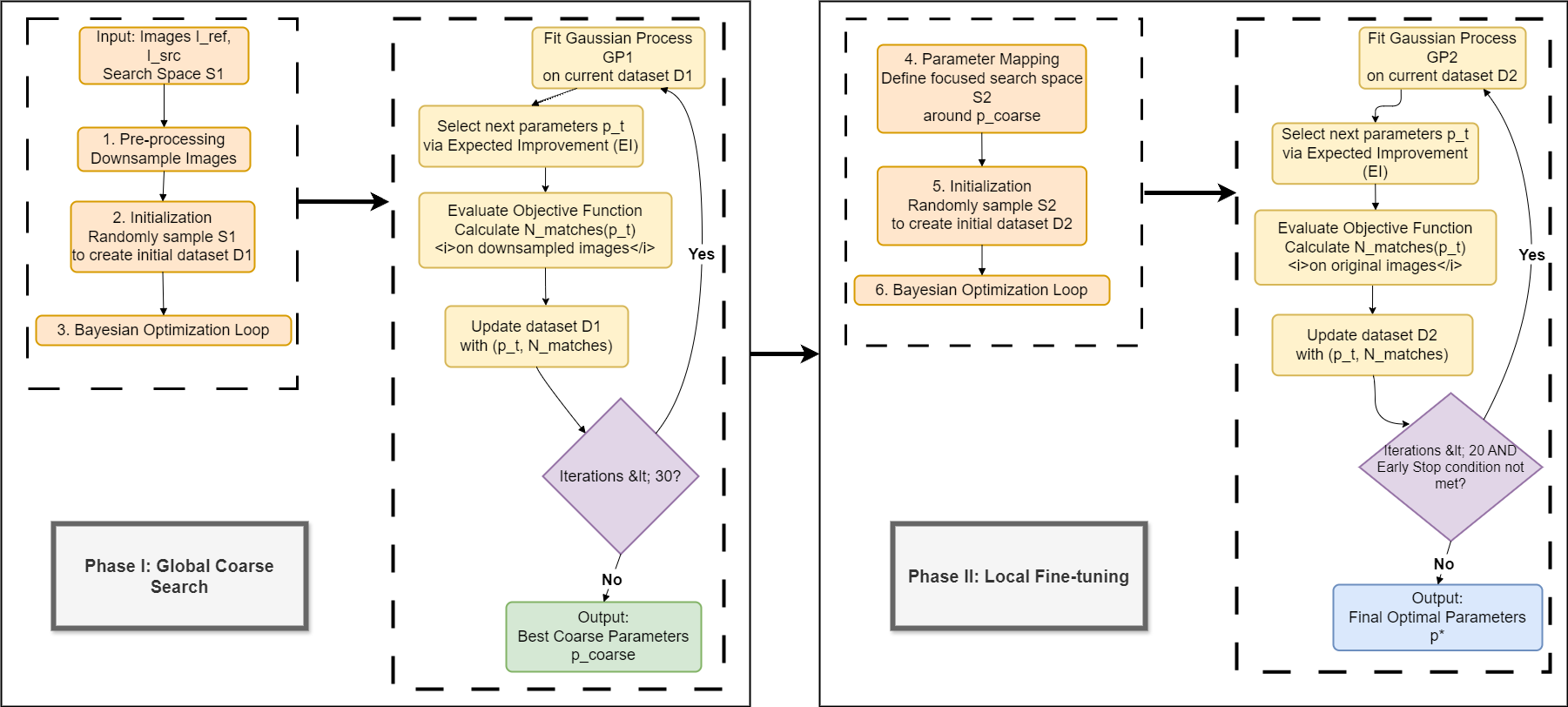}
		\caption{The complete workflow of our HBO framework. The process begins with Phase I (Global Coarse Search) on downsampled images to identify promising parameter regions, followed by Phase II (Local Fine-tuning) on full-resolution images to determine the optimal solution.}
		\label{fig:methodology_flowchart}
	\end{figure*}

	Figure \ref{fig:methodology_flowchart} illustrates the complete workflow of our HBO framework. It is a two-phase approach that synergistically balances computational speed with optimization accuracy to intelligently navigate the parameter space.
	
	The first phase consists of a Global Coarse Search performed on downsampled images. The core rationale behind this step is that the overall topology of the objective function's response surface is largely preserved even at a lower resolution. This assumption is firmly rooted in scale-space theory \citep{Lindeberg1994}, which demonstrates that downsampling effectively acts as a low-pass filter, suppressing high-frequency noise while preserving the essential structural information of the stellar profiles. According to the principle of scale-space causality, the local extrema of the objective function—representing the optimal parameter regions—evolve continuously across different resolutions rather than appearing or shifting sporadically. Furthermore, as the alignment success depends on the relative geometric configurations of point sources, these patterns remain fundamentally scale-invariant \citep{Lowe2004}, ensuring that high-performance regions identified in the coarse space accurately represent the global optimum's neighborhood. Consequently, we can dramatically decrease the computational cost per evaluation, enabling the BO algorithm to explore a wide, global parameter space rapidly and affordably. This allows the framework to efficiently discard vast, unpromising regions and identify a high-potential "hyper-volume" where the optimal parameters are most likely to reside.
	
	Following the coarse search, the framework transitions to the second phase: a Local Fine-tuning Search. This stage operates on the original, full-resolution images, ensuring that the final parameters are optimized for the highest-fidelity data. The search space for this phase is no longer global; instead, it is a smaller, localized region centered around the best-performing parameters found in Phase I. This focused approach allows the BO algorithm to dedicate its computational budget to meticulously refining the solution, achieving sub-pixel precision without the risk of being trapped in a suboptimal local minimum far from the true global optimum. In essence, this hierarchical strategy leverages the strengths of both scales: the speed of low-resolution exploration to find the right "neighborhood" and the precision of high-resolution exploitation to find the exact "address."
	
	\subsection{Defining the optimization problem: parameter space and objective function}
	\label{sec:problem_definition}
	
	The core of our adaptive framework is to reframe the manual task of parameter tuning as a formal optimization problem. This requires a precise definition of the search space (the parameters to be optimized) and an objective function (the metric to be maximized). Our design choices are guided by the need to create a problem landscape that is both representative of alignment quality and efficiently navigable by the Bayesian Optimization algorithm.
	
	\subsubsection{Parameter space design}
	The primary drivers of source detection performance in algorithms like DAOStarFinder are FWHM and SNR threshold. Therefore, our optimization is focused on the two-dimensional parameter vector $p = \{\text{FWHM}, \text{SNR}\}$.
	
	The rationale for selecting these parameters is their direct physical significance. FWHM quantifies the stellar profile's size, which is a function of atmospheric seeing, telescope optics, and focus. A precise FWHM value enables the algorithm's kernel to accurately model the star's shape, crucial for distinguishing real sources from artifacts. The SNR threshold acts as a filter, determining the minimum confidence level required to accept a detection. It creates a critical trade-off: a low threshold may increase the number of true, faint sources but at the cost of introducing a flood of spurious noise detections, while a high threshold yields a clean but potentially sparse source list.
	
	Defining an appropriate search space for these parameters is vital for the efficiency of the BO algorithm. The space must be broad enough to accommodate a wide range of observational conditions yet compact enough to avoid wasting evaluations in physically implausible regions. Our hierarchical strategy uses two distinct search spaces, $S_1$ for the global coarse search and $S_2$ for the local fine-tuning. The primary space, $S_1$, is designed as follows (summarized in Table \ref{tab:core_params}):
	
	   FWHM Search Range: The range is defined as [1.0, 8.0] pixels, a selection specifically calibrated to the optical specifications of the 1-meter telescope employed in this study. Given the instrument's pixel scale of 0.21$^{\prime\prime}$/pixel, this search interval corresponds to a physical angular seeing range of approximately 0.21$^{\prime\prime}$ to 1.68$^{\prime\prime}$. This range is strategically chosen to encompass the full spectrum of atmospheric conditions at the observation site, from rare near-diffraction-limited nights to more common poor seeing conditions. By anchoring the absolute pixel bounds to the physical sampling rate of the specific detector and telescope optics, we ensure that the search space remains physically relevant and robust for data acquired across different observational epochs.
	
	   SNR Search Range: The lower bound is fixed at 3.0, a statistically motivated threshold corresponding to a 3$\sigma$ confidence level, which effectively filters out most random background noise. The upper bound, however, is not a fixed value but is adaptively determined as $\text{SNR}_{\text{max}}$. This value is calculated from the brightest, non-saturated source in the reference image. This adaptive approach is a key design choice: it automatically tailors the search space to the quality of the specific dataset, preventing the optimizer from exploring excessively high SNR values that are irrelevant for a given noisy image.
	
	In addition to FWHM and SNR, source detection algorithms often rely on morphological parameters such as `sharpness` and `roundness` to further filter spurious detections.
	Sharpness measures the acutance of a source, indicating how point-like an object is. It is typically defined as:
	\begin{equation}
		\text{sharpness} = \frac{\text{peak} - \text{mean(sky)}}{\text{FWHM}},
	\end{equation}
	where $\text{peak}$ is the peak brightness of the source and $\text{mean(sky)}$ is the local background level. For a well-formed stellar object, sharpness values typically fall within a specific range, e.g., $[0.2, 1.0]$.
	Roundness, on the other hand, quantifies the symmetry of a source profile, helping to distinguish stars from cosmic rays or elongated galaxies:
	\begin{equation}
		\text{roundness} = 2 \times \frac{\text{FWHM}_x - \text{FWHM}_y}{\text{FWHM}_x + \text{FWHM}_y}.
	\end{equation}
	Here, $\text{FWHM}_x$ and $\text{FWHM}_y$ are the full widths at half maximum along the major and minor axes, respectively. For perfectly symmetric stars, roundness approaches zero, while it ranges from $[-1.0, 1.0]$ for elongated or unresolved objects.
	
	To ensure the objective function's landscape remains smooth and continuous, which is beneficial for Bayesian Optimization, we adopt a robust design for these ancillary parameters. Instead of imposing restrictive, "hard" thresholds (e.g., sharpness in $[0.2, 1.0]$) that can lead to abrupt and undesirable drops in the number of detected sources with minor parameter changes, we deliberately relax their filtering criteria to their maximum possible ranges (i.e., sharpness $[0.0, 1.0]$ and roundness $[-1.0, 1.0]$). The justification for this relaxation is twofold: First, it avoids creating a "spiky" or discontinuous objective function surface that could mislead the optimizer. Second, it shifts the responsibility of rejecting true outliers (such as cosmic rays, hot pixels, or elongated galaxies) from the initial source detection stage to the subsequent geometric transformation step, which employs the highly robust Random Sample Consensus (RANSAC) algorithm \citep{fischler1981}. RANSAC is inherently designed to identify and exclude extreme outliers during the matching process, ensuring that only genuinely matched and spatially coherent source pairs contribute to the final transformation, even if the initial source list contains some false positives. This approach ensures that the Bayesian optimizer can efficiently explore the primary FWHM-SNR space without being prematurely constrained by overly strict morphological filters.
	
	\begin{table}[ht]
		\caption{Hierarchical search space for optimization parameters.}
		\label{tab:core_params}
		\centering
		\setlength{\tabcolsep}{5pt}
		\begin{tabular*}{\columnwidth}{@{\extracolsep{\fill}} l l l}
			\hline\hline
			Parameter & Phase I: Global Search & Phase II: Local Tuning \\
			\hline
			FWHM & [1.0, 8.0] & $p_{\text{coarse, FWHM}} \pm \Delta_{\text{FWHM}}$ \\
			SNR  & [3.0, $\text{SNR}_{\text{max}}$] & $p_{\text{coarse, SNR}} \pm \Delta_{\text{SNR}}$ \\
			\hline
		\end{tabular*}
		\tablefoot{
			The search space for Phase II is dynamically centered around the best point found in Phase I, $p_{\text{coarse}} = \{p_{\text{coarse, FWHM}}, p_{\text{coarse, SNR}}\}$. The search windows, $\Delta_{\text{FWHM}}$ and $\Delta_{\text{SNR}}$, are typically set to a fraction of the global range (e.g., 20\%) to ensure a focused yet comprehensive local search. $\text{SNR}_{\text{max}}$ is adaptively determined from the reference image.
		}
	\end{table}
	
	This adaptive search space design directly solves the problem of "static parameters in dynamic conditions" mentioned in the introduction. By calculating $\text{SNR}_{\text{max}}$ in real-time, the algorithm automatically adjusts its sensitivity based on the specific noise floor and source brightness of each frame. This ensures that the search remains focused on physically relevant regions, preventing the optimizer from wasting computational effort on thresholds that would be either too shallow for deep images or too aggressive for noisy ones.
	
	\subsubsection{Objective function formulation}
	
	Having defined the parameter space, the next critical step is to establish an objective function, $f(p)$, that quantitatively evaluates the quality of an alignment produced by a given parameter set $p = \{\text{FWHM}, \text{SNR}\}$. The ideal objective function should be a reliable proxy for the final alignment success and precision, while also being computationally efficient and well-behaved for the optimization algorithm.
	
	Our primary goal in image alignment is to derive a robust geometric transformation that accurately maps one image onto another. The reliability of this transformation, typically computed using algorithms like RANSAC, is directly contingent on the number of confident, spatially consistent source pairs—or "inliers"—that can be identified between the two images. A larger number of matched pairs provides a stronger statistical basis for the transformation, making it more resilient to noise and spurious detections. Therefore, we select the number of successfully matched source pairs, denoted as $N_{\text{matches}}$, as the core metric for our objective function.
	
	This choice is motivated by several key advantages. First, $N_{\text{matches}}$ is a direct and intuitive measure of alignment quality. If an insufficient number of pairs is found (e.g., fewer than the minimum required to solve for a transformation, typically 3 for an affine transform), the alignment fails outright. As the number increases, the solution becomes more stable and robust. Second, this metric synergizes perfectly with our strategy of relaxing the `sharpness` and `roundness` filters. The underlying matching tool, \texttt{Astroalign} \citep{Beroiz2020}, employs a RANSAC-based algorithm that is inherently designed to sift through a potentially noisy list of sources and identify the largest subset that conforms to a single geometric transformation. Consequently, maximizing $N_{\text{matches}}$ implicitly encourages parameter combinations that produce a rich set of true sources while relying on RANSAC to handle the outliers, thus directly optimizing for a successful geometric solution.
	
	Standard Bayesian Optimization libraries are typically designed to solve minimization problems. To align with this convention, we formulate our objective function as the negative of the number of matched pairs. The optimization problem is thus defined as finding the parameter set $p^*$ that minimizes this function:
	\begin{equation}
		f(p) = -N_{\text{matches}}(p),
	\end{equation}
	
	This objective function is specifically designed to navigate the two failure modes discussed in the introduction. An "overly strict" parameter set will yield a sparse source list and thus a low $N_{\text{matches}}$, while an "overly lenient" set will introduce excessive noise that causes the matching algorithm to fail the consensus check, also resulting in a low $N_{\text{matches}}$. By maximizing this single value, the Bayesian optimizer is naturally guided to the "sweet spot"—a configuration that is sensitive enough to find sufficient true sources but selective enough to maintain a clean list for the RANSAC-based geometric matching.
	
	such that the optimal parameter set $p^*$ is given by:
	\begin{equation}
		p^* = \arg\min_{p \in S} f(p) = \arg\max_{p \in S} N_{\text{matches}}(p).
	\end{equation}
	In this framework, each evaluation of the objective function for a given $p$ involves: (1) detecting sources in both the source and reference images using the candidate FWHM and SNR values, and (2) feeding these source lists into the matching algorithm to compute $N_{\text{matches}}$. The result, a single scalar value, is then returned to the Bayesian optimizer, which uses it to update its internal model of the parameter landscape.
	
	\subsection{Bayesian optimization framework}
	
	With the optimization problem formally defined, we now detail the engine driving the search: BO. Unlike grid search or random search, which sample the parameter space without memory, BO is an intelligent, sequential strategy that builds a probabilistic model of the objective function and uses this model to decide where to sample next. Following standard practical Bayesian optimization practices \citep{Snoek2012}, this approach is particularly well-suited for problems where the objective function is expensive to evaluate, as is the case with our image alignment task. The BO framework consists of two core components: a probabilistic surrogate model to represent the objective function, and an acquisition function to guide the search. 
	
	To replace the subjectivity and lack of guarantee for global optimality in manual tuning, we employ the BO engine. Unlike manual adjustments, which are often local and biased by the user's experience, BO provides a mathematically rigorous and objective strategy to explore the parameter landscape, ensuring that the selected FWHM and SNR are the globally optimal solutions for the given dataset.
	
	\subsubsection{Probabilistic surrogate model: gaussian process}
	\label{sec:gp_model}
	
	The first component of BO is a surrogate model that approximates the unknown, expensive-to-evaluate objective function $f(p)$. We employ a Gaussian Process for this purpose, a powerful and flexible non-parametric model widely used in Bayesian inference. The key advantage of a GP is its ability to provide not only a prediction for the objective function's value at a new point, but also a measure of uncertainty about that prediction. This uncertainty is crucial for balancing exploration and exploitation.
	
	A Gaussian Process defines a distribution over functions. We can think of it as a generalization of a multivariate Gaussian distribution to an infinite-dimensional function space. A function $f(p)$ is said to follow a GP if any finite collection of points $\{p_1, ..., p_n\}$ from its domain has a corresponding set of function values $\{f(p_1), ..., f(p_n)\}$ that follows a multivariate Gaussian distribution. A GP is fully specified by its mean function $\mu(p)$ and its covariance function (or kernel) $k(p, p')$, which models the correlation between the function values at two different points, $p$ and $p'$. We can write this as:
	\begin{equation}
		f(p) \sim \mathcal{GP}(\mu(p), k(p, p')),
	\end{equation}
	where $\mu(p)$ represents our prior belief about the mean of the function (often set to zero for simplicity), and $k(p, p')$ defines the smoothness and other properties of the function.
	
	The choice of the kernel is critical as it encodes our assumptions about the function we are modeling. For our problem, we use the squared exponential kernel (also known as the Radial Basis Function or RBF kernel), a common and effective choice for smooth functions:
	\begin{equation}
		k(p, p') = \sigma_f^2 \exp \left( -\frac{1}{2} \sum_{i=1}^{d} \frac{(p_i - p'_i)^2}{\ell_i^2} \right).
	\end{equation}
	Here, $p$ and $p'$ are $d$-dimensional parameter vectors. The hyperparameter $\sigma_f^2$ is the signal variance, controlling the overall vertical variation of the function, while each $\ell_i$ is a length-scale parameter that determines how quickly the correlation between points decays with distance along the $i$-th dimension. These hyperparameters are typically learned from the data by maximizing the marginal likelihood. The rationale for using this kernel is that we expect the number of matched sources to vary smoothly with small changes in FWHM and SNR, rather than exhibiting erratic, high-frequency oscillations.
	
	The BO process begins with a few initial evaluations of the true objective function $f(p)$ at randomly selected points, forming an initial dataset $D_t = \{(p_1, y_1), ..., (p_t, y_t)\}$, where $y_i = f(p_i)$. Given this data, we can update our GP prior to obtain a posterior distribution over functions. For any new candidate point $p_*$, this posterior distribution is also a Gaussian, $P(f(p_*) | D_t, p_*) = \mathcal{N}(\mu_t(p_*), \sigma_t^2(p_*))$, with a predictive mean $\mu_t(p_*)$ and a predictive variance $\sigma_t^2(p_*)$. The mean $\mu_t(p_*)$ represents the model's best guess for the function's value at $p_*$, while the variance $\sigma_t^2(p_*)$ quantifies the uncertainty of that guess. The variance will be low near points we have already evaluated and high in unexplored regions of the parameter space. This interplay between the predicted value and its uncertainty is precisely what the acquisition function will leverage to make intelligent decisions.
	
	\subsubsection{Decision strategy: the acquisition function}
	\label{sec:acquisition_func}
	
	Once the GP surrogate model has been updated with the available data to produce a posterior distribution over the objective function, the next step is to decide which point in the parameter space, $p$, should be evaluated next. This decision is governed by an acquisition function, denoted as $\alpha(p)$. The acquisition function's role is to quantify the "utility" or "desirability" of sampling any given point. The point that maximizes the acquisition function is chosen as the next candidate for evaluation:
	\begin{equation}
		p_{t+1} = \arg\max_{p \in S} \alpha(p | D_t).
	\end{equation}
	The acquisition function is constructed using the GP's predictive mean $\mu_t(p)$ and variance $\sigma_t^2(p)$, and its design is key to navigating the trade-off between exploitation and exploration. Exploitation means sampling in regions where the surrogate model predicts a low objective value (i.e., a high number of matches), while exploration involves sampling in regions of high uncertainty, where the true value of the objective function might be even better than what has been observed so far.
	
	Several acquisition functions exist, such as Probability of Improvement (PI) and Upper Confidence Bound (UCB). For our framework, we have chosen the Expected Improvement acquisition function, one of the most widely used and effective strategies. The rationale for this choice is its excellent performance in practice and its principled approach to balancing the exploration-exploitation dilemma. EI quantifies the expected amount of improvement over the best value found so far, $y_{\text{best}} = \min(y_1, ..., y_t)$.
	
	For any candidate point $p$, we first define the improvement function $I(p)$ as the difference between the current best value and the (unknown) function value at $p$, ensuring it is non-negative:
	\begin{equation}
		I(p) = \max(0, y_{\text{best}} - f(p)).
	\end{equation}
	Since $f(p)$ is a random variable according to our GP model (specifically, $f(p) \sim \mathcal{N}(\mu_t(p), \sigma_t^2(p))$), the improvement $I(p)$ is also a random variable. The EI acquisition function is simply the expectation of this improvement, taken over the GP posterior distribution:
	\begin{equation}
		\alpha_{\text{EI}}(p) = \mathbb{E}[I(p) | D_t].
	\end{equation}
	This expectation has a convenient closed-form analytical solution, which makes its computation highly efficient:
	\begin{equation}
		\alpha_{\text{EI}}(p) = (y_{\text{best}} - \mu_t(p))\Phi(Z) + \sigma_t(p)\phi(Z),
	\end{equation}
	where $Z = \frac{y_{\text{best}} - \mu_t(p)}{\sigma_t(p)}$, and $\Phi(\cdot)$ and $\phi(\cdot)$ are the cumulative distribution function (CDF) and probability density function (PDF) of the standard normal distribution, respectively.
	
	The structure of the EI function elegantly balances the two competing objectives. The first term, $(y_{\text{best}} - \mu_t(p))\Phi(Z)$, favors points where the predicted mean is significantly lower than the current best value (exploitation). The second term, $\sigma_t(p)\phi(Z)$, favors points where the model's uncertainty is high (exploration). EI will be large either in regions where the model is confident of a good outcome, or in regions where the model is very uncertain, potentially hiding an even better outcome. By selecting the point that maximizes EI at each step, our algorithm systematically and efficiently converges towards the global optimum of the objective function, adaptively focusing its search efforts on the most promising areas of the parameter space.
	
	The synergy between the GP surrogate model and the EI acquisition function allows the framework to make "informed decisions" about which parameters to test next. This significantly reduces the total number of evaluations required to find the optimal solution compared to an exhaustive grid search, providing the necessary efficiency for the automated processing of massive astronomical data in large-scale sky surveys.
	
	\begin{algorithm}[ht!]
		\caption{HBO for image alignment}
		\label{alg:hbo}
		\begin{algorithmic}
			\Require
			Reference image $I_{\text{ref}}$, Source image $I_{\text{src}}$
			\Require
			Global search space $S_1$, Local search space rule $S_{2, \text{rule}}$
			\Require
			Iterations $N_1=30$ (coarse), $N_2=20$ (fine)
			\Require
			Initial samples $N_{\text{init1}}=10$, $N_{\text{init2}}=5$
			\Require
			Early stopping patience $T=5$
			\Ensure
			Optimal parameters $p^*$, Maximum matches $N_{\text{matches}}^*$
			
			\Statex \Comment{Define the objective function for convenience}
			\Function{EvaluateObjective}{$p, I_a, I_b$}
			\State \Return $-N_{\text{matches}}(p, I_a, I_b)$
			\EndFunction
			\Statex
			
			\State Phase I: Global Coarse Search on Downsampled Images
			\State $I_{\text{ref\_down}}, I_{\text{src\_down}} \gets \text{Downsample}(I_{\text{ref}}, I_{\text{src}}, \text{factor}=2)$
			\State $D_1 \gets \emptyset$ \Comment{Initialize coarse search dataset}
			\State Select initial points $\{p_{1,1}, \dots, p_{1,N_{\text{init1}}}\}$ randomly from $S_1$
			\For{$i = 1 \to N_{\text{init1}}$}
			\State $y_{1,i} \gets \text{EvaluateObjective}(p_{1,i}, I_{\text{ref\_down}}, I_{\text{src\_down}})$
			\State $D_1 \gets D_1 \cup \{(p_{1,i}, y_{1,i})\}$
			\EndFor
			\For{$t = N_{\text{init1}} + 1 \to N_1$}
			\State Fit Gaussian Process model $GP_1$ on $D_1$
			\State $p_{1,t} \gets \operatornamewithlimits{argmax}_{p \in S_1} \alpha_{\text{EI}}(p | GP_1)$
			\State $y_{1,t} \gets \text{EvaluateObjective}(p_{1,t}, I_{\text{ref\_down}}, I_{\text{src\_down}})$
			\State $D_1 \gets D_1 \cup \{(p_{1,t}, y_{1,t})\}$
			\EndFor
			\State $p_{\text{coarse}} \gets \operatornamewithlimits{argmin}_{(p,y) \in D_1} y$ \Comment{Best parameters from coarse search}
			\Statex
			
			\State Phase II: Local Fine-tuning on Original Images
			\State $p_{\text{coarse}}.\text{FWHM} \gets p_{\text{coarse}}.\text{FWHM} \times 2$ \Comment{Scale FWHM for original resolution}
			\State Define local search space $S_2$ based on $S_{2, \text{rule}}$ around $p_{\text{coarse}}$
			\State $D_2 \gets \emptyset$; $\text{count} \gets 0$; $y_{\text{best}} \gets +\infty$
			\State Select initial points $\{p_{2,1}, \dots, p_{2,N_{\text{init2}}}\}$ randomly from $S_2$
			\For{$i = 1 \to N_{\text{init2}}$}
			\State $y_{2,i} \gets \text{EvaluateObjective}(p_{2,i}, I_{\text{ref}}, I_{\text{src}})$
			\State $D_2 \gets D_2 \cup \{(p_{2,i}, y_{2,i})\}$
			\If{$y_{2,i} < y_{\text{best}}$}
			\State $y_{\text{best}} \gets y_{2,i}$; $\text{count} \gets 0$
			\EndIf
			\EndFor
			\For{$t = N_{\text{init2}} + 1 \to N_2$}
			\If{$\text{count} \ge T$}
			\State break \Comment{Early stopping triggered}
			\EndIf
			\State Fit Gaussian Process model $GP_2$ on $D_2$
			\State $p_{2,t} \gets \operatornamewithlimits{argmax}_{p \in S_2} \alpha_{\text{EI}}(p | GP_2)$
			\State $y_{2,t} \gets \text{EvaluateObjective}(p_{2,t}, I_{\text{ref}}, I_{\text{src}})$
			\State $D_2 \gets D_2 \cup \{(p_{2,t}, y_{2,t})\}$
			\If{$y_{2,t} < y_{\text{best}}$}
			\State $y_{\text{best}} \gets y_{2,t}$; $\text{count} \gets 0$
			\Else
			\State $\text{count} \gets \text{count} + 1$
			\EndIf
			\EndFor
			\State $p^* \gets \operatornamewithlimits{argmin}_{(p,y) \in D_2} y$
			\State $N_{\text{matches}}^* \gets -y_{\text{best}}$
			\State \Return $p^*, N_{\text{matches}}^*$
		\end{algorithmic}
	\end{algorithm}
	
	\section{Experiments and results}
	\label{sec:results}
	
	The experimental evaluation of the HBO framework is structured to rigorously address the core limitations of traditional astronomical image alignment identified in Section \ref{sec:intro}. Each experiment is designed as a direct response to a specific problem: the failure of static parameters in fluctuating environments, the subjective and inefficient nature of manual tuning, and the necessity for sub-pixel precision in automated pipelines. 
	
	To rigorously evaluate the performance of our method, we utilized high-fidelity observational data from the 1-meter telescope at the Yunnan Astronomical Observatory. The technical specifications of this instrument are summarized in Table \ref{tab:telescope_specs}. In the following subsections, we present a detailed diagnostic analysis of these experiments and provide a physical interpretation of the results.
	
	\begin{table}[htbp]
		\centering
		\caption{Technical specifications of the 1-meter telescope at Yunnan Observatory.}
		\label{tab:telescope_specs}
		\begin{tabular*}{\columnwidth}{@{\extracolsep{\fill}}ll}
			\hline\hline 
			Property & Value \\
			\hline
			Aperture (Primary mirror) & 101.6 cm \\
			Focal length & 1330 cm \\
			F-ratio & 13 \\
			CCD pixel array & 2048 $\times$ 2048 \\
			Field of View (FOV) & 7.1$^{\prime}$ $\times$ 7.1$^{\prime}$ \\
			Pixel scale & 0.21$^{\prime\prime}$/pixel \\
			Site Location & IAU code 286 \\
			\hline
		\end{tabular*}
		\tablefoot{These optical parameters determine the image scale and are critical for interpreting the FWHM values optimized by our framework.}
	\end{table}
	
	Our experiments focus on two distinct near-Earth objects (NEOs): 66391 Moshup and 01036 Ganymed.
	
	The dataset for 66391 Moshup was captured over three consecutive nights from June 2 to June 4, 2019. This sequence includes 30 CCD observation images. All frames were obtained using an I-band filter to ensure consistent spectral response. Each individual image was taken with a fixed exposure time of 15 seconds.
	
	Similarly, the observations for 01036 Ganymed were conducted between June 2 and June 4, 2019. This dataset comprises 33 CCD images. These observations also utilized the I-band filter for high-quality detection of faint targets. Following the same protocol as Moshup, each frame for Ganymed was captured with a single exposure time of 15 seconds. These diverse observational sequences provide a robust testbed for evaluating the adaptability of our Bayesian optimization framework under varying astronomical conditions.
	
	\subsection{Contrast test: from default failure to HBO-driven success}
	\label{sec:contrast_test}
	
	The first experiment is a direct diagnostic of the "static parameter failure" problem. As argued in the introduction, rigid thresholds often lead to "Overly Strict" detection failures when observing conditions deteriorate. We utilize the NEO 66391 Moshup dataset, captured under poor seeing conditions, to illustrate this phenomenon.
	
	Initially, we applied the standard "default" parameters ($p_{\text{fixed}} = \{\text{FWHM}=3.5, \text{SNR}=5.0\}$), which are typical for automated pipelines in median-seeing conditions. The results were catastrophic: the alignment failed completely because the stellar profiles in the image were significantly blurred beyond 3.5 pixels. The star-finding algorithm, mismatching the PSF with its narrow kernel, effectively ignored the real stars and treated them as background noise. This resulted in an empty or critically sparse source list, providing no geometric basis for alignment.
	
	To solve this, we employed the HBO framework on the same dataset. Unlike the static approach, HBO autonomously recognized the blurred state of the stars and identified the optimal parameters: $p^* = \{\text{FWHM}=12.83, \text{SNR}=5.37\}$. By expanding the FWHM kernel to match the actual atmospheric seeing of 12.8 pixels, the algorithm successfully "recovered" the lost stellar signals. 
	
\begin{figure*}[ht] 
		\centering
		\includegraphics[width=0.9\textwidth]{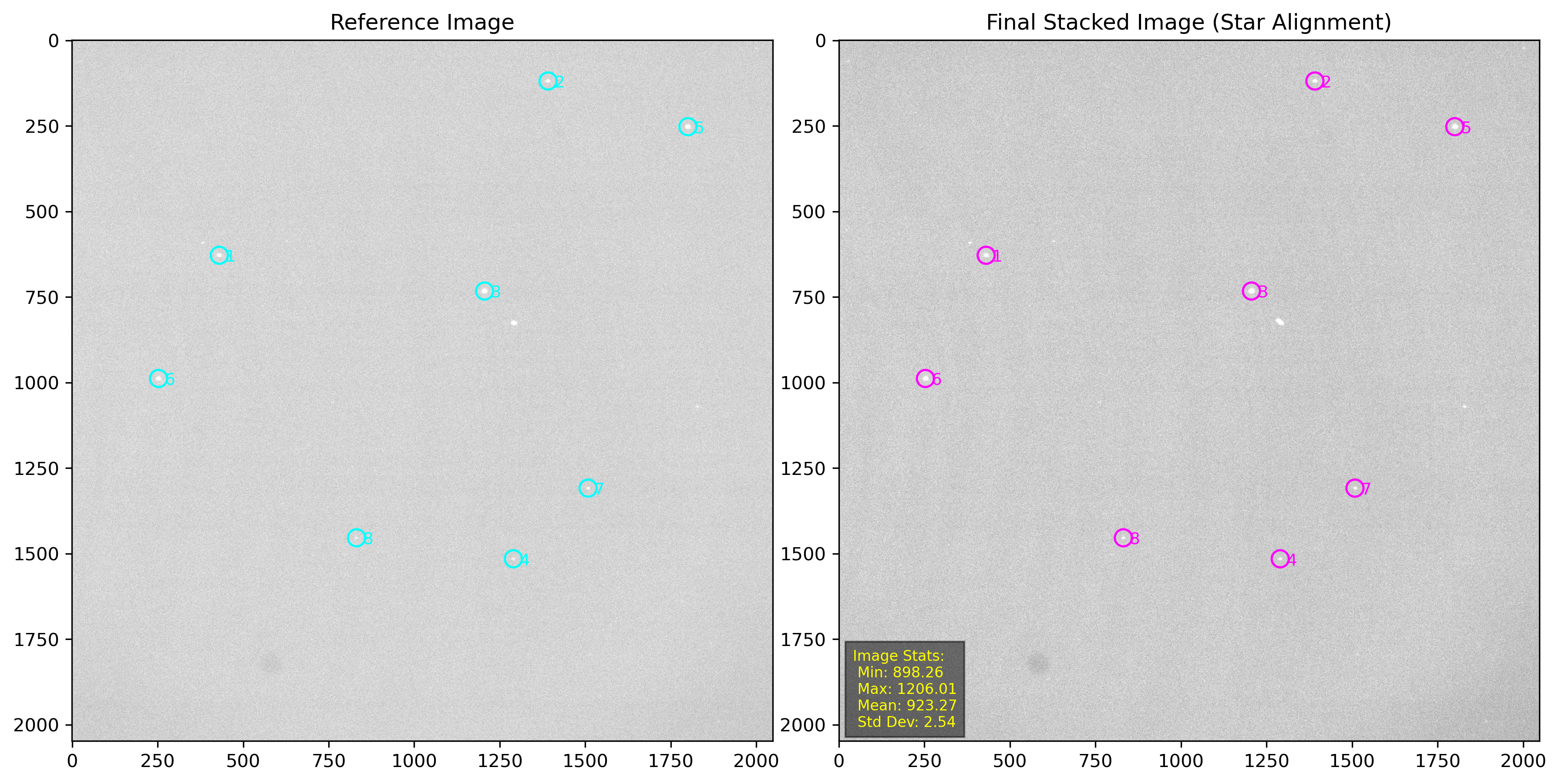}
		\caption{The ultimate success of the HBO framework on the Moshup dataset where traditional methods failed. The left panel highlights the 8 successfully} identified and matched source pairs (the "geometric consensus"). The right panel shows the final, high-precision stacked image, proving that the adaptive search successfully resolved the seeing-mismatch problem and achieved sub-pixel alignment stability.
		\label{fig:result_experimental}
	\end{figure*}

	The visual proof of this success is showcased in Fig. \ref{fig:result_experimental}. The left panel illustrates the "constellation" of successfully matched source pairs. These 8 pairs represent the geometric bridge that the HBO framework built between the noisy images—a bridge that the fixed method was unable to construct. The right panel shows the final stacked image. Every star is perfectly circular and sharp, and the signal-to-noise ratio of the faint object has been significantly improved. This contrast provides a powerful justification for our work: it demonstrates that our adaptive framework can transform a failed observation into a scientifically valid result by simply "re-tuning" the detection logic to match the physical reality of the night.
	
	\subsection{Mapping the landscape: the objective logic of bayesian inference}
	\label{sec:bo_landscape}
	
	Having demonstrated that HBO can solve the alignment failure, we now explore the underlying logic of the search itself. A major problem with manual tuning is its "subjectivity and lack of global optimality." To address this, we must prove that the HBO search is not a lucky guess but a systematic reconstruction of the parameter-response landscape. 
	
\begin{figure*}[ht]
		\centering
		\includegraphics[width=\textwidth]{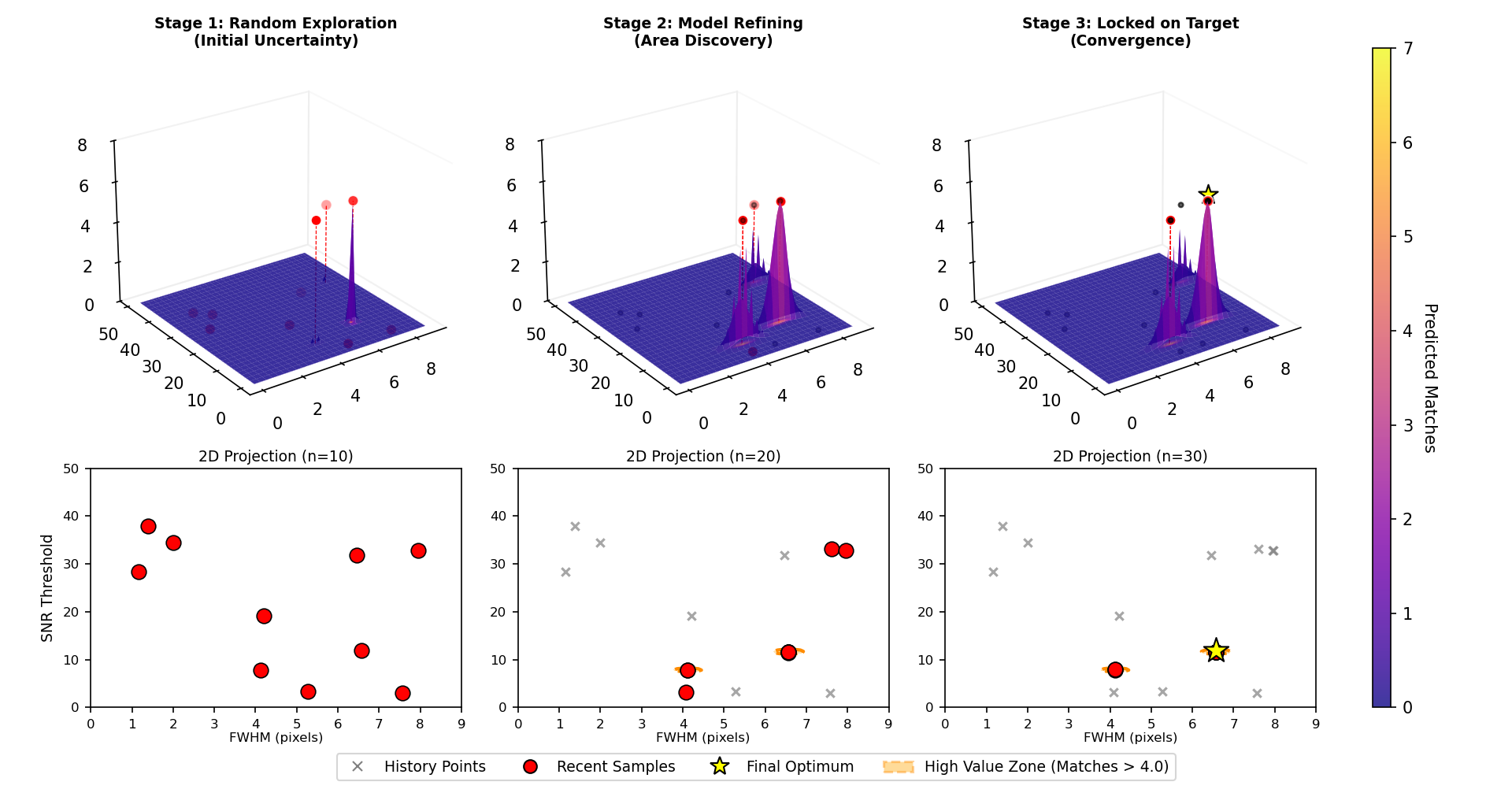} 
		\caption{Spatio-temporal evolution of the GP surrogate model for the Moshup dataset. These three stages represent the transition from total uncertainty (Stage 1) to the discovery of high-performance regions (Stage 2) and final convergence on the global optimum (Stage 3). This process ensures that the chosen parameters are objectively the best for the given observational data.}
		\label{fig:hbo_evolution}
	\end{figure*}

	Fig. \ref{fig:hbo_evolution} visualizes the evolution of the GP surrogate model during the 30-iteration search. This figure explains the internal "decision-making" process of the algorithm:
	\begin{itemize}
		\item Stage 1 (n=10 trials): Initial Exploration. This stage represents the algorithm's state of maximal uncertainty. The 3D surface is mostly flat, meaning the model has not yet identified any correlation between parameters and success. The samples are scattered randomly to provide a global "skeleton" of the space.
		\item Stage 2 (n=20 trials): Topographical Discovery. After 20 evaluations, the GP model has begun to "learn" the topography. We see yellow peaks emerging in the high-FWHM region. The 2D projection below shows the formation of orange "High Value Zones." The algorithm has objectively narrowed down the search to the region where the FWHM kernel correctly models the physical stars.
		\item Stage 3 (n=30 trials): Global Convergence. The model has now reconstructed a well-defined peak, and the final samples are concentrated at the gold star (the global optimum). This visualization proves that our framework replaces human intuition with a mathematically rigorous, probabilistic inference of the observing conditions.
	\end{itemize}
	
	\subsection{Scientific validity: convergence of astrometric precision}
	\label{sec:rms_convergence}
	
\begin{figure}[h!]
		\centering
		\includegraphics[width=0.9\hsize]{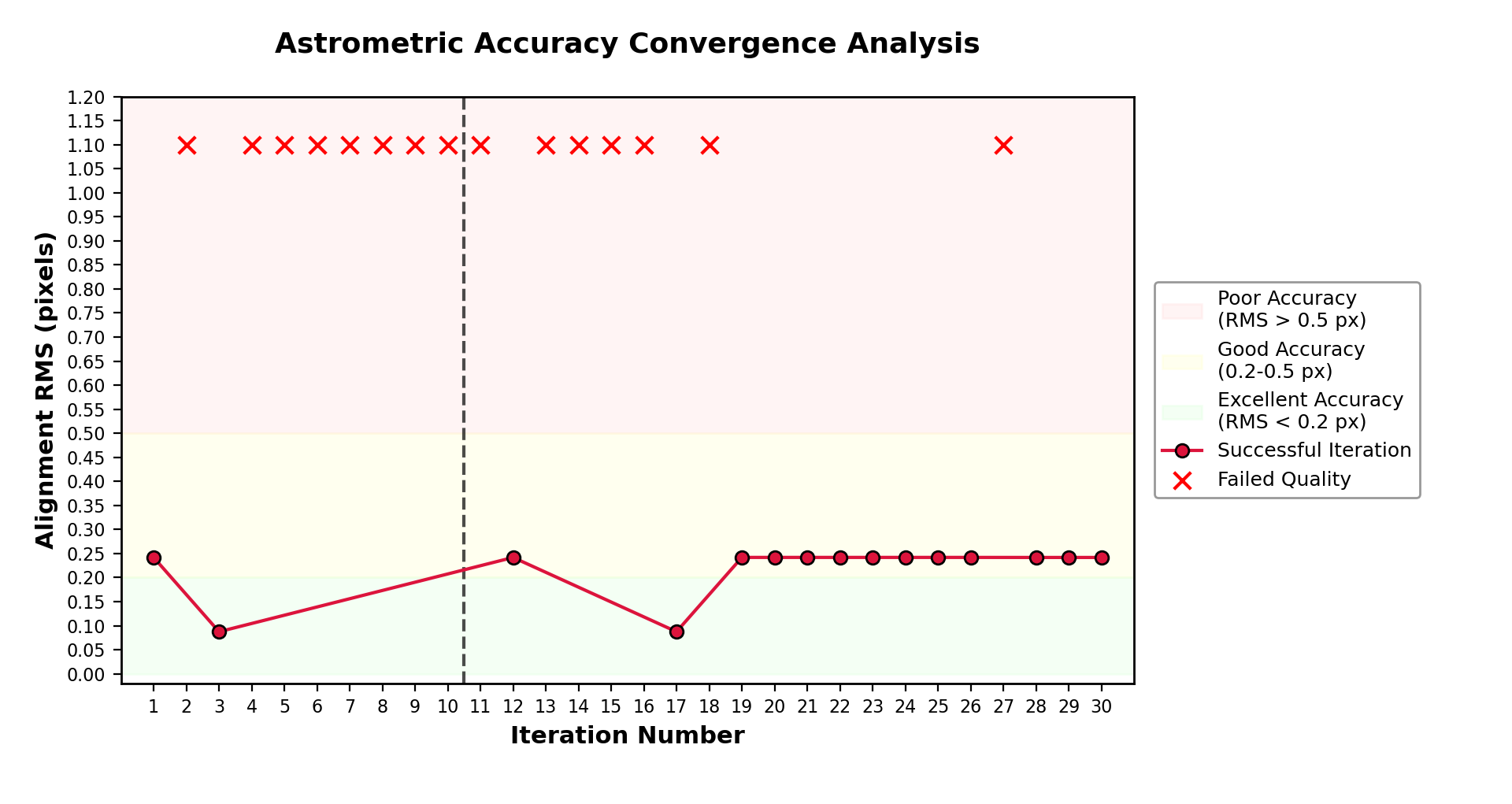}
		\caption{Real-time convergence of astronomical image alignment precision during the search. The synchronized drop in RMS residuals (circles) as the algorithm discovers more matches confirms that $N_{\text{matches}}$ is an exceptionally effective proxy for the physical quality of the alignment.}
		\label{fig:rms_fluctuation}
	\end{figure}

	Obtaining high-precision centroid data of celestial objects serves as the foundation for achieving high-accuracy astrometric results, which can be realized through image alignment and stacking. The precision of alignment and stacking is positively correlated with the centroiding accuracy of celestial objects in the stacked images. A key concern in automated pipelines is whether the optimization of a proxy metric ($N_{\text{matches}}$) leads to the goal: high-precision centroiding. Therefore, to prove this link, we analyze the real-time convergence of the astronomical image alignment precision using RMS residuals in Fig. \ref{fig:rms_fluctuation}.
	
	The search dynamics shown in Fig. \ref{fig:rms_fluctuation} are highly telling. In the first half of the search (iterations 1-15), the RMS error is erratic and often hits the "Failure" ceiling (red crosses), because the algorithm is still testing inappropriate parameters. However, as the Bayesian engine identifies parameter combinations that increase $N_{\text{matches}}$, the RMS error undergoes a synchronized and sharp decline. 
	
	The best precision of 0.0874 pixels achieved in the green "Excellent Accuracy" band is a significant result. It proves that our choice of objective function—maximizing the number of matches—is scientifically sound. In the presence of centroiding noise, a larger number of matched stars provides a more "over-determined" geometric solution, which suppresses individual errors. This justifies the fundamental architecture of our framework: quantity of matches leads directly to the quality of registration.
	
	\subsection{Efficiency and trajectory analysis for automated surveys}
	\label{sec:efficiency_analysis}
	
\begin{figure}[h!]
		\centering
		\includegraphics[width=0.95\columnwidth]{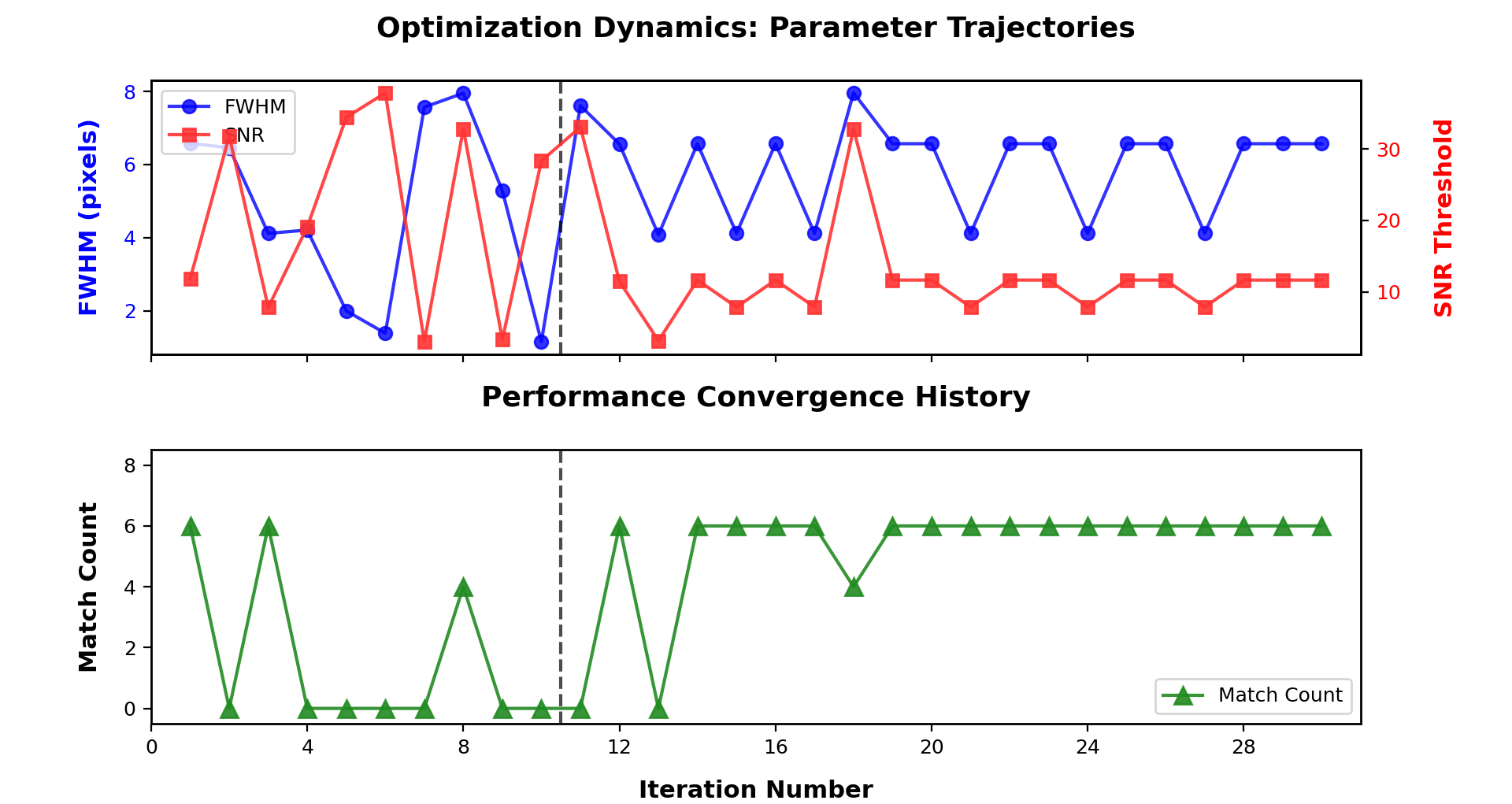}
		\caption{Trajectories of FWHM and SNR optimization over 30 iterations. The rapid stabilization of the parameters (top) and the corresponding increase in match counts (bottom) highlight the efficiency and convergence reliability of the HBO framework.}
		\label{fig:optimization_trajectory}
	\end{figure}

	To address the "inefficiency of manual trial-and-error" mentioned in the introduction, we must prove that HBO is fast enough for massive data processing. The total optimization time for Moshup was 47.19 seconds, which we analyze through the parameter trajectories in Fig. \ref{fig:optimization_trajectory}.
	
	The upper panel of Fig. \ref{fig:optimization_trajectory} shows that after an initial "zigzag" phase of exploration, both the FWHM and SNR trajectories flatten into stable lines by iteration 25. This rapid stabilization is a direct result of our hierarchical strategy: by using downsampled images for the initial phase, we prune the search space at a low computational cost, allowing the full-resolution fine-tuning to reach a solution quickly. This speed ensures that our method can be integrated into real-time sky survey pipelines without becoming a bottleneck.
	
	\subsection{Robustness test: inter-night stability and variable-seeing sequences}
	\label{sec:robustness_test}
	
	The final experiment addresses the "high failure rate" problem under complex, real-world conditions by analyzing image sequences of a single target captured over multiple nights. From a physical perspective, multi-night observations are essential for the high-precision astrometry and orbit determination of Near-Earth Objects (NEOs), such as Moshup and Ganymed, where a long time baseline is required to predict future trajectories. However, such data is inherently heterogeneous: each night presents a unique "atmospheric realization" of turbulence, transparency, and sky brightness. Traditionally, this heterogeneity required a human expert to manually re-tune the pipeline for every observing run to avoid PSF-mismatch. Our goal here is to evaluate the HBO framework's ability to act as a self-correcting interface, ensuring that long-term monitoring can be conducted with full automation and consistent registration quality.
	
	We conducted an inter-night stability test across three different observing nights for the Moshup dataset (Table \ref{tab:robustness_conditions}) and an intra-night fluctuation test on the Ganymed sequence. While the baseline fixed-parameter method failed on 100\% of the nights due to its rigid thresholds, the HBO framework maintained a 100\% success rate. Even on the night of 2019-06-03, which was characterized by exceptionally poor seeing and a heavily blurred PSF, HBO recovered a successful alignment with an RMS of 0.124 pixels. By adaptively determining the optimal FWHM for each specific epoch, HBO ensures that the final registration maintains a homogeneous quality tier despite the physical variability of the raw data. This ability to "re-sense" the state of the night ensures that the scientific integrity of the long-term baseline is preserved without human intervention.
	
\begin{figure*}[ht]
		\centering
		\includegraphics[width=0.9\textwidth]{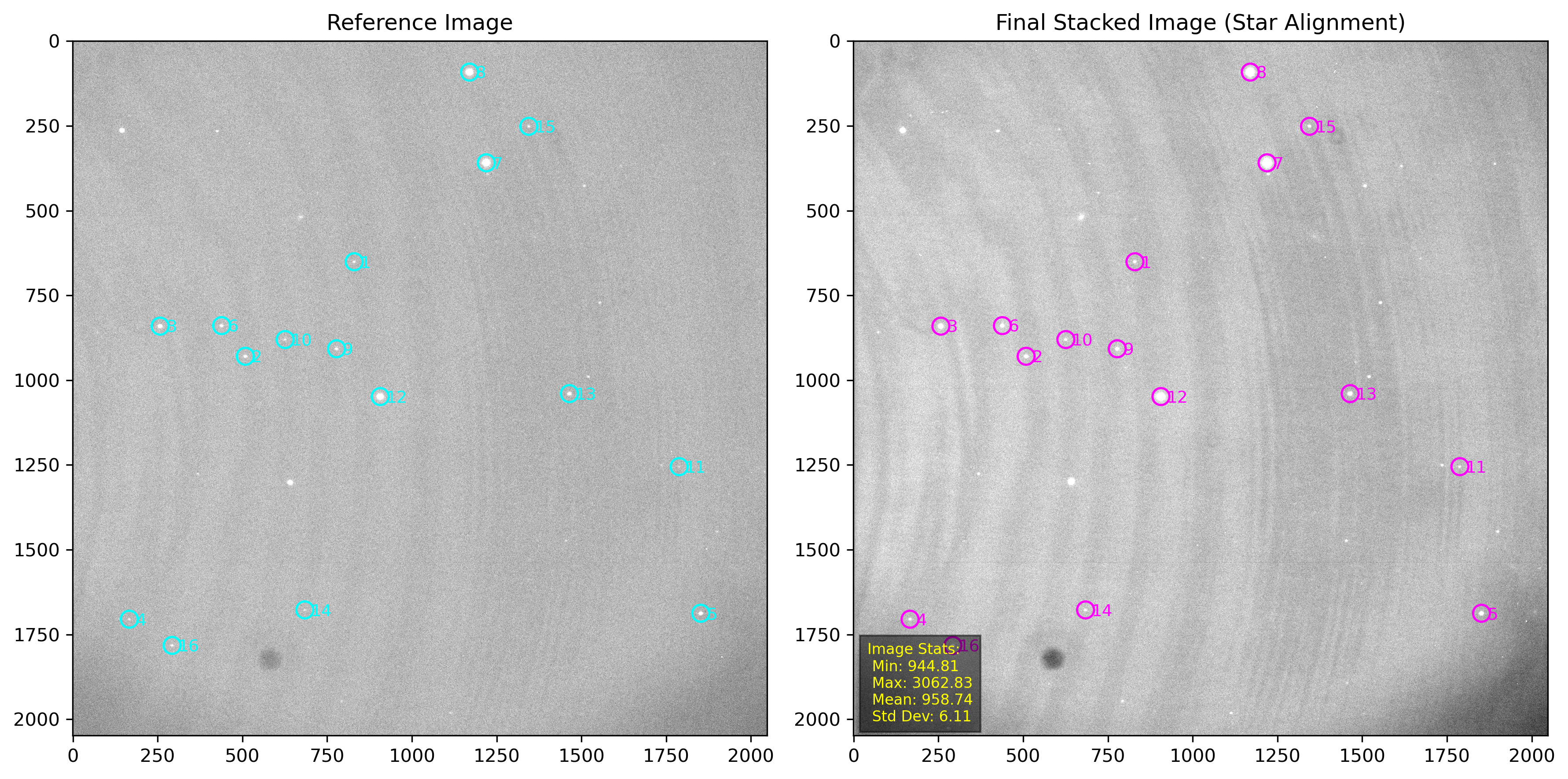}
		\caption{Alignment result for the Ganymed dataset sequence. The framework's frame-to-frame adaptation ensures that stacking remains precise and clear even when the observing conditions vary rapidly within a single observational run.}
		\label{fig:result_ganymed_adaptive}
	\end{figure*}

	The framework's adaptability was further confirmed using the Ganymed sequence, where image quality fluctuated rapidly from frame to frame due to atmospheric jitter. As shown in Fig. \ref{fig:result_ganymed_adaptive}, HBO's ability to adjust its detection kernel on-the-fly ensured that the final stack remained sharp and free of the "smearing" artifacts that typically occur when alignment is not perfectly matched to the nightly PSF. This visual confirmation, combined with the 100\% success rate across diverse conditions, proves that the HBO framework is a robust, efficient, and objective solution. The proposed adaptive strategy demonstrates the potential to mitigate the high failure rates associated with static parameter configurations, offering a more flexible approach for future automated astronomical data processing pipelines.\textsf{}
	
	\begin{table}[htbp] 
		\centering
		\caption{Comparative robustness analysis for Moshup across three observing nights.}
		\label{tab:robustness_conditions}
		\begin{tabular*}{\columnwidth}{@{\extracolsep{\fill}}lccc}
			\hline\hline
			Date & Baseline & HBO & Best RMS \\
			\hline
			2019-06-02 & Failure & Success ($N=8$) & 0.087 \\
			2019-06-03 & Failure & Success ($N=8$) & 0.124 \\
			2019-06-04 & Failure & Success ($N=10$)& 0.091 \\
			\hline
		\end{tabular*}
		\tablefoot{The baseline failure is a direct result of PSF-mismatch, a problem that the HBO framework solves by adaptively re-tuning the star-detection kernel for each specific night.}
	\end{table}
	
	\section{Conclusions}
	\label{sec:conclusions}
	
	In this study, we have explored an adaptive parameter optimization approach for astronomical image alignment, focusing on the automated selection of source detection parameters, specifically FWHM and the SNR threshold. By implementing a hierarchical Bayesian optimization framework, we sought to minimize the reliance on manual trial-and-error in the data processing workflow. The two-phase search strategy developed here offers a practical means of navigating the parameter space, attempting to balance optimization accuracy with computational requirements through the use of downsampled images for global exploration.
	
	The performance of the HBO framework was evaluated using observational data of the near-Earth objects 66391 Moshup and 01036 Ganymed, obtained from the 1-meter telescope at the Yunnan Astronomical Observatory in June 2019. These specific datasets, captured under varying seeing conditions, provided a baseline for testing the framework's adaptability. Our results indicate that the adaptive strategy can identify functional alignment parameters across different observing nights. In the cases analyzed, the method effectively mitigated registration failures caused by static-parameter mismatches. By adjusting the FWHM kernel to match the actual atmospheric conditions, the framework facilitated successful matching even in instances where fixed-threshold approaches yielded insufficient source pairs.
	
	Quantitative evaluation of the Moshup sequence showed that the optimization process reached a minimum astrometric RMS residual of 0.0874 pixels. Regarding computational cost, the total optimization time for this dataset was recorded at 47.19 seconds in our experimental setup, with parameter stabilization typically observed within approximately 25 iterations. These metrics suggest that the HBO framework can enhance the automation level of image alignment pipelines for the specific conditions encountered in our test cases.
	
	Although these initial results demonstrate the feasibility of the proposed adaptive approach, the scope of this study is limited to a finite set of observational scenarios and targets. The robustness of the framework in more complex environments, such as extremely crowded fields, has not yet been fully investigated. Future work will focus on refining the objective function and exploring additional morphological constraints to improve the reliability of source extraction. Furthermore, the application of more sophisticated probabilistic models, including Gaussian Cox Processes, may be necessary to better account for non-Gaussian background noise. Extensive testing on larger and more diverse datasets remains essential to determine the generalizability and long-term reliability of this method for large-scale astronomical surveys.
	
	\begin{acknowledgements}
		We acknowledge the support from the staff at the 1 m telescope administered in Yunnan Observatories. This research work is financially supported by the National Natural Science Foundation of China (grant Nos. 12173085) and the Training Object Project of technological innovation talents in Yunnan Province(No. 202305AD160004). 
	\end{acknowledgements}
	
	\bibliographystyle{aa} 
	\bibliography{references} 
	
\end{document}